\documentclass[twocolumn,prb,showpacs]{revtex4}

\usepackage{graphicx}
\usepackage{rotating}
\usepackage{amsmath}
\usepackage{amsfonts}
\usepackage{amssymb}
\usepackage{enumerate}
\usepackage{longtable}
\usepackage{dcolumn}
\usepackage{bbm}
\usepackage{psfrag}

\begin{document}

\title{Hole Dispersions for Antiferromagnetic Spin-$\frac{1}{2}$ 
Two-Leg Ladders by Self-Similar Continuous Unitary Transformations}

\author{S. Duffe}
\affiliation{Lehrstuhl f\"{u}r Theoretische Physik I, 
Technische Universit\"{a}t Dortmund,
 Otto-Hahn Stra\ss{}e 4, 44221 Dortmund, Germany}

\author{G. S. Uhrig}
\affiliation{Lehrstuhl f\"{u}r Theoretische Physik I, 
Technische Universit\"{a}t Dortmund,
 Otto-Hahn Stra\ss{}e 4, 44221 Dortmund, Germany}

\date{\rm\today}

\begin{abstract}
The hole-doped antiferromagnetic spin-$\frac{1}{2}$ two-leg ladder is an 
important model system for the high-$T_c$ superconductors based on cuprates. 
Using the technique of self-similar continuous unitary transformations we 
derive effective Hamiltonians for the charge motion in these ladders. The key 
advantage of this technique is that it provides effective models explicitly
in the thermodynamic limit. A real space restriction of the 
generator of the transformation allows us to explore the experimentally 
relevant parameter space. From the effective Hamiltonians we calculate the 
dispersions for single holes. Further calculations will enable the calculation 
of the interaction of two holes so that a handle of Cooper pair formation is 
within reach.
\end{abstract}

\pacs{74.72.Gh, 71.10.Fd, 71.15.Qe, 75.10.Kt}


\maketitle

\section{Introduction}
\label{chap:intro}
The hole-doped antiferromagnetic spin-$\frac{1}{2}$ two-leg ladder is used as 
a model system, theoretically as well as experimentally, for the 
two-dimensional cuprate 
superconductors. Because the spin ladder is quasi one-dimensional it can more
easily be treated by  numerical and analytical approaches than the
full two-dimensional model. For this reason the 
spin ladder was subject to many theoretical  investigations in recent years 
\cite{dagot88,dagot92b,barne93,dagot96,troye96,white97,jecke98,oitma99,sushk99,johns00a,brunn01,jurec01,jurec02,schmi05b,roux05,bouil11}. From a strong 
coupling perspective it is reasonable to describe the doped spin ladder by a 
$t$-$J$-model \cite{harri67,zhang88}. 

In this paper, our aim is to derive an effective Hamiltonian
which describes not only the motion of the magnetic degrees of
freedom, i.e., the triplons \cite{schmi03c},
 but also of the charges, i.e., of the doped holes. This task
is more challenging than the description of the triplons alone
because the two kinds of excitations interact strongly and 
their energy ranges are not separated but they strongly overlap. Hence
the excitations are not infinitely long-lived but they decay, at least
in certain regions of the Brillouin zone.

The technique employed here is the approach of
self-similar continuous unitary transformations 
(SCUT). In this approach the mapping of the true ground state to a
vacuum of elementary excitations is constructed systematically,
details will be explained in Sect.\ \ref{chap:SCUT}. 
The basic idea is to implement a transformation 
that adjusts itself during the procedure of the diagonalisation depending on 
the current form of the Hamiltonian at this instant of the continuous 
transformation. The change of the Hamiltonian induced by the transformation is 
determined by the current magnitude of the non-diagonal elements.
The flowing Hamiltonian is given in second quantization, i.e.,
it has a certain structure in terms of elementary creation and
annihilation operators \cite{knett03a}. This formulation
incorporates the linked cluster property \cite{oitma06} automatically.
The SCUT systematically defines a set
of differential equations for the prefactors of the monomials of
the creation and annihilation operators.

The key issue is to construct the flow to the effective Hamiltonian
in a robust way. Due to the energetically overlapping states this goal required
modifications \cite{fisch10a} 
with respect to the previously performed unitary transformations.
Thereby we achieved the successful calculation of the dispersion
of a single hole.

In Sects.\ \ref{chap:hole disp without ring exc} and 
\ref{chap:hole disp with ring exc} we investigate the dispersions of single 
hole excitations obtained from the effective Hamiltonians, which are results 
of the SCUT. We also compare our results to the results of other methods.

Sects.\ \ref{chap:summ} and \ref{chap:outlook} conclude this article with a 
summary and perspectives for future work.

\section{Spin-$\frac{1}{2}$ Ladders}
\label{chap:spin ladders}
\begin{figure}
\begin{center}
\psfrag{u}{u}
\psfrag{l}{l}
\psfrag{ri}{$n$}
\psfrag{ri1}{$n+1$}
\psfrag{Js}{$J_{\perp}, t_{\perp}$}
\psfrag{Jp}{$J_{\parallel}, t_{\parallel}$}
\psfrag{Jc}{$J_{\square}$}
\includegraphics[scale=0.6]{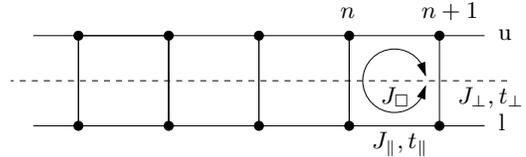}
\caption{Spin-$\frac{1}{2}$ ladder consisting of the upper leg u, the lower leg
 l and the rungs $n$. The magnetic couplings are the nearest neighbor coupling 
on the rungs $J_{\perp}$, the nearest neighbor coupling along the legs 
$J_{\parallel}$ and the four spin coupling $J_{\square}$ for the ring exchange.
 The nearest neighbor hopping is described by the constants $t_{\perp}$ (on the
 rungs) and $t_{\parallel}$ (along the legs). The symmetry axis of the ladder 
is depicted by the dashed line.}
\label{ladder}
\end{center}
\end{figure}
The spin-$\frac{1}{2}$ ladder is realised in the so called telephone number 
compounds $\mathrm{(Sr, La, Ca, Y)}_{14}\mathrm{Cu}_{24}\mathrm{O}_{41}$ 
\cite{vulet06}. For these systems superconductivity can be detected under 
high pressure \cite{uehar96}. Interactions between the ladders are weak 
because they result from $90^{\circ}$ exchange \cite{gopal94, matsu00b}. Thus 
the ladders can be considered to be  isolated from one other. Even if the 
interladder coupling is taken into account, the strong frustration of the 
lattice causes the system to be effectively one-dimensional \cite{schmi07a}. 
The dispersions of the magnetic excitations of the complete layer with 
interladder coupling are similar to those of a single spin ladder. Important 
magnetic properties of the ladders are investigated in Refs.\ 
\onlinecite{dagot88, dagot92b, barne93, dagot96, johns00a, schmi05b}.

The focus of the present article is the lightly doped spin ladder because
this will allow us to address the dynamics of charges. Of course,
this system has been investigated previously by many techniques,
for instance exact diagonalization \cite{troye96,roux05}, 
density-matrix renormalization \cite{white97,jecke98,roux05}, 
high-order series expansion \cite{oitma99},
diagrammatic approaches \cite{sushk99,jurec01,jurec02},  
and quantum Monte Carlo \cite{brunn01}. Hence much is known about
the dynamics of a single hole, i.e., its dispersion,
its spectral density and to some extent also about the 
bound states of two holes. 
The aim of the present paper is to establish an approach which
generates an effective model for the motion of holes and their interaction
with magnetic excitations and other holes. Such a model can be expressed
in terms of creation and annihilation operators which create and annihilate
holes, i.e., charge excitations, and triplons, i.e., magnetic 
excitations. The techniques used so far do not yield such an effective
model but numerical data for eigen energies and spectral weights.
For this reason we here use the complementary approach of 
self-similar continuous unitary transformations.

The spin sites in our model (see Fig.\ \ref{ladder}) are the $3d_{x^2-y^2}$ 
orbitals of the copper atoms coupled via the $2p_x$ or $2p_y$ orbitals of the 
oxygen atoms, which hybridise with the copper $3d_{x^2-y^2}$ orbitals so that 
an antiferromagnetic superexchange \cite{ander50} is possible.
The Hamiltonian in the usual form of a $t$-$J$-model, for derivations
see Ref.\ \onlinecite{hamer10} and references therein, reads
\begin{eqnarray}
\label{Hamiltonian}
H &=& J_{\perp} \sum_{n} \vec{S}_{n, \mathrm{u}} \cdot \vec{S}_{n, \mathrm{l}} 
+ J_{\parallel} \sum_{n, \alpha=\mathrm{u,l}} \vec{S}_{n, \alpha} \cdot 
\vec{S}_{n + 1, \alpha} 
\\
&& + J_{\square} \sum_n \left[ \left( \vec{S}_{n, \mathrm{u}} \cdot 
\vec{S}_{n+1, \mathrm{u}} \right) \left( \vec{S}_{n, \mathrm{l}} \cdot 
\vec{S}_{n+1, \mathrm{l}} \right) \right. 
\nonumber\\
&& + \left( \vec{S}_{n, \mathrm{u}} \cdot \vec{S}_{n, \mathrm{l}} \right) 
\left( \vec{S}_{n+1, \mathrm{u}} \cdot \vec{S}_{n+1, \mathrm{l}} \right) 
\nonumber\\
&& \left. - \left( \vec{S}_{n, \mathrm{u}} \cdot \vec{S}_{n+1, \mathrm{l}} 
\right) \left( \vec{S}_{n, \mathrm{l}} \cdot \vec{S}_{n+1, \mathrm{u}} \right) 
\right] 
\nonumber\\
&& - t_{\perp} \sum_{n, \alpha=\mathrm{u,l}} 
c^{\dagger}_{\sigma, \alpha, n} 
c^{\phantom{\dagger}}_{\sigma, \overline{\alpha}, n} - 
t_{\parallel} \sum_{n, \alpha=\mathrm{u,l}} c^{\dagger}_{\sigma, \alpha, n} 
c^{\phantom{\dagger}}_{\sigma, \alpha, n\pm1}.
\nonumber 
\end{eqnarray}
The nearest neighbor terms with the couplings $J_{\perp}$ and $J_{\parallel}$ 
(see Fig.\ \ref{ladder}) are not sufficient to describe the magnetic 
interactions in the system. Because the hybridization path around the 
$\mathrm{Cu}_4\mathrm{O}_4$ square plaquettes is strong, the influence of the 
four-spin interactions belonging to these plaquettes (i.e.\ two neighboring 
rungs) is not negligible \cite{colde01b}. These interactions are referred to as
 ring exchange (also cyclic exchange) with the coupling $J_{\square}$ (see 
Fig.\ \ref{ladder}). Actually the complete ring exchange also includes two-spin
 terms for all two-spin combinations of the four spins on two neighboring 
rungs. Yet the contributions from the terms coupling the spins along the rungs 
and parallel to the legs are merged with the nearest neighbor terms so that 
the coupling constants $J_{\perp}$ and $J_{\parallel}$ include these 
contributions, whereas the terms coupling the spins diagonally can be 
neglected because their prefactor is only of the order of 
$3\%$ of $J_{\perp}$ \cite{mizun99}. The complete representation of the ring 
exchange using spin operators can be found in Ref.\ \onlinecite{brehm99}.

The hopping of the spins is described by the constants $t_{\perp}$ and 
$t_{\parallel}$ (see Fig.\ \ref{ladder}). The creation and annihilation 
operators $c^{\dagger}$ and $c^{\phantom{\dagger}}$ already incorporate 
hardcore properties so that double occupancy is forbidden.\par
We define the dimensionless quantities
\begin{equation}
x=\frac{J_{\parallel}}{J_{\perp}},\,x_{\square}=\frac{J_{\square}}{J_{\perp}},
\,\lambda_{\perp}=\frac{t_{\perp}}{J_{\perp}},\,\lambda_{\parallel}
=\frac{t_{\parallel}}{J_{\perp}},
\end{equation}
which are normalized to $J_{\perp}$.\par
It is convenient to treat the ladder as a one-dimensional chain of rungs. 
There are nine possible local states on a rung $n$ of the ladder.
\begin{subequations}
\begin{eqnarray}
& & \label{singlet}\left|s\right\rangle_{n} = \frac{1}{\sqrt{2}} 
\left( \left|\uparrow\downarrow\right\rangle - 
\left|\downarrow\uparrow\right\rangle \right)_{n}
\\
& & \label{tripletx}\left|t_x\right\rangle_{n} = \frac{-1}{\sqrt{2}} 
\left( \left|\uparrow\uparrow\right\rangle - 
\left|\downarrow\downarrow\right\rangle \right)_{n}
\\
& & \label{triplety}\left|t_y\right\rangle_{n} = \frac{\mathrm{i}}{\sqrt{2}} 
\left( \left|\uparrow\uparrow\right\rangle + 
\left|\downarrow\downarrow\right\rangle \right)_{n}
\\
& & \label{tripletz}\left|t_z\right\rangle_{n} = \frac{1}{\sqrt{2}} 
\left( \left|\uparrow\downarrow\right\rangle + 
\left|\downarrow\uparrow\right\rangle \right)_{n}
\\
& & \label{hole1}\left|a_{\tau = 1, \sigma = 1}\right\rangle_{n} = 
\frac{1}{\sqrt{2}} \left( \left|\uparrow0\right\rangle + 
\left|0\uparrow\right\rangle \right)_{n}
\\
& & \label{hole2}\left|a_{\tau = -1, \sigma = 1}\right\rangle_{n} = 
\frac{1}{\sqrt{2}} \left( \left|\uparrow0\right\rangle - 
\left|0\uparrow\right\rangle \right)_{n}
\\
& & \label{hole3}\left|a_{\tau = 1, \sigma = -1}\right\rangle_{n} = 
\frac{1}{\sqrt{2}} \left( \left|\downarrow0\right\rangle + 
\left|0\downarrow\right\rangle \right)_{n}
\\
& & \label{hole4}\left|a_{\tau = -1, \sigma = -1}\right\rangle_{n} = 
\frac{1}{\sqrt{2}} \left( \left|\downarrow0\right\rangle - 
\left|0\downarrow\right\rangle \right)_{n}
\\
& & \label{2holes}\left|2h\right\rangle_{n} = \left|00\right\rangle_{n}
\end{eqnarray}
\end{subequations}
The states without holes are the singlet (\ref{singlet}) and the triplet 
states (\ref{tripletx}-\ref{tripletz}), for which we use the representation 
from Ref. \onlinecite{sachd90} because the symmetry in spin space is manifest 
in this representation. The states consisting of one hole and one spin 
$\frac{1}{2}$ (\ref{hole1}-\ref{hole4}) have the quantum numbers $\tau$, which 
is the parity of the state with respect to the symmetry axis of the ladder 
(see dashed line in 
Fig.\ \ref{ladder}), and $\sigma$, which indicates whether the $S_z$ 
component of the spin is up or down. The double hole state (\ref{2holes}) is 
neglected here because only one-hole states will be considered. The local 
energy of $\left|2h\right\rangle$ is also larger than the energy of the other 
states. Therefore $\left|2h\right\rangle$ is expected to be also 
less important for ladders with low, but macroscopic doping, though
it will influence the quantitative results for hole-hole scattering
and hole-hole binding.

For the application of the continuous unitary transformation a representation 
using the creation and annihilation operators for the local states 
(\ref{singlet}-\ref{hole4}) is indicated. The complete Hamiltonian 
(\ref{Hamiltonian}) in the respresentation using these operators can be found 
in Ref.\ \onlinecite{duffe10}. Note that the local rung states incorporate 
hardcore properties. Also the fermionic states exhibit additional hardcore 
properties which simply
result from the fact that each rung can only be in {\it one} particular state.

The excitations are actually dominated by the local states as long as the 
couplings between the rungs are small. The magnetic excitations are triplons, 
i.e., triplet states dressed with the magnetic interactions with their 
environment \cite{schmi03c}. We do not want to use the term ``magnon'' for 
these magnetic excitations because this term is usually associated with 
quasiparticles in systems that exhibit long-range magnetic order (which is not 
the case for the spin ladder). Moreover triplons feature the
appropriate  threefold degeneracy
 based on their $S=1$ character \cite{schmi03c}.

\section{SCUT}
\label{chap:SCUT}
We use the method of continuous unitary transformations (CUT) to obtain an 
effective Hamiltonian \cite{glaze93,wegne94,kehre06}. 
Instead of applying only a single 
constant unitary transformation that diagonalises the Hamiltonian at once or 
several constant unitary transformations successively, a unitary transformation
 depending on a continuous parameter is applied to the Hamiltonian. This 
transformation adjusts itself permanently during its application. Discrete 
transformations must be known explicitly before we can apply them, whereas for 
the continuous transformation it is sufficient to set up the infinitesimal 
generator of the transformation. The unitary operator $U$ depending on the 
continuous parameter $l$ transforms the Hamiltonian by
\begin{equation}
\label{transformation}
H(l) = U(l) H U^{\dagger}(l).
\end{equation}
The generator of the transformation is defined by
\begin{equation}
\label{generator}
\eta(l) = \frac{\partial U(l)}{\partial l} U^{\dagger}(l).
\end{equation}
The generator defines the properties of the transformation. The derivative of 
$H$ with respect to $l$ is given by the so-called flow equation
\begin{equation}
\label{flow equation}
\frac{\partial H(l)}{\partial l} = \left[ \eta(l), H(l)\right]\,\mathrm{.}
\end{equation}
This is actually a system of differential equations 
(generically highly coupled) for the coefficients of the operators appearing 
in $H(l)$. To achieve the effective Hamiltonian the flow equation has to be 
solved.
For a self-similar application of the CUT \cite{mielk97b,reisc04,fisch10a}
the Hamiltonian is represented by a 
sum of different operators $\hat{o}_i$, which are products of local operators 
in second quantization, multiplied by prefactors $g_i$.
\begin{equation}
H(l) = \sum_i g_i(l) \hat{o}_i
\end{equation}
The operators $\hat{o}_i$ determine the structure of $H(l)$. 
This is the reason why this formulation  of the CUTs
 is called self-similar. In particular, we express
the operators $\hat{o}_i$ in terms of creation and annihilation
operators of charge excitations (holes) and  spin excitations
(triplons) in the manner introduced previously \cite{knett03a}.
A reference state, the vacuum, has to be chosen with respect to which these
creation and annihilation processes are defined.
In our case, this state is the product of local singlets on
the rungs of the ladder. For the undoped ladder this
state is a convenient starting point as seen in
many high-order perturbative \cite{uhrig96b,trebs00,knett01b} and
diagrammatic approaches \cite{jurec01,jurec02}.
Since we focus in this work on single hole dynamics
the use of the product of local singlets on the rungs
as reference vacuum is well justified.

The differential equations for the 
$g_i(l)$ are given by the flow equation in the form
\begin{equation}
\label{flow equation for coefficients}
\frac{\partial g_i(l)}{\partial l} = \sum_{j, k} a_{i, j, k} g_j(l) g_k(l).
\end{equation}
Because an infinite system yields an infinite number of differential equations
which cannot be solved numerically,
 we have to apply a truncation to $H(l)$ so that a closed system of 
differential equations is obtained. We choose a real space truncation because 
the correlation length is finite for the spin ladder due to the energy gap 
\cite{dagot88, dagot92b}. The concrete truncation scheme is explained in 
Sect.\ \ref{chap:trunc_scheme}.

It should also be noted that the truncation is only restricted to operators. 
There is no truncation of the Hilbert space of the states. Thus although 
operators affecting higher numbers of particles are omitted, the number of 
particles that can be treated is arbitrarily large. To illustrate this we 
consider the action of a term in second quantisation, e.g., 
$a^{\dagger}_i a^{\dagger}_j a^{\phantom{\dagger}}_k$, which annihilates one 
particle and creates two. This term acts not only on the one-particle subspace,
 but also on the subspaces with more than one particle, e.g.\ it changes a 
four-particle into a five-particle state.

The solution of the flow equation is a rather straightforward numerical 
integration. The convergence is monitored during the integration. Since all 
contributions to the generator decrease in case of convergence, the absolute 
values of the concerning coefficients are squared and summed. The square root 
of this sum is defined as residual off-diagonality (ROD). It is the norm of the
 generator and also a measure for the convergence. The ROD is expected to tend 
to zero for $l\rightarrow\infty$. Note that the term ``off-diagonality'' is 
meant in a broader sense, i.e.\ the definition of the generator determines 
which elements shall be kept for $l\rightarrow\infty$ and these elements are 
defined as ``diagonal'' parts of the Hamiltonian. The RODs depicted in the 
following are always normalized to the initial ROD at $l=0$.

With the decrease of the ROD the wanted effective Hamiltonian is 
approached. If the relative ROD falls below a certain threshold specifying the 
precision of the result (usually $\approx10^{-8}$), the integration can be 
considered to be completed. The number of coefficients treated in our
actual calculations is of the order up to $10^5$ for the doped spin ladder.

\subsection{Generator}
\label{chap:gen}

\begin{figure}
\begin{center}
\psfrag{l=0}{$l = 0$}
\psfrag{finite l}{finite $l$}
\psfrag{l=inf}{$l = \infty$}
\includegraphics[scale=0.4]{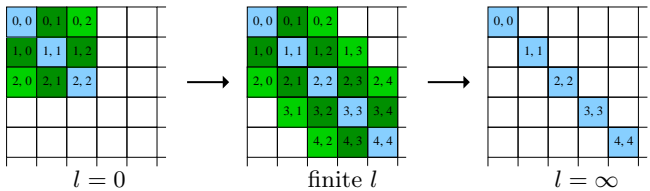}
\caption{(Color online) Schematic example for a transformation of $H(l)$ 
induced by the pc generator: Each coloured ($n_c$, $n_a$)-block represents 
contributions to $H$ creating $n_c$ particles after annihilating $n_a$ 
particles.}
\label{pc CUT blocks}
\end{center}
\end{figure}

The particle conserving or pc generator
\begin{equation}
\label{pc generator}
\eta_{\mathrm{pc}, i, j}(l) = \mathrm{sign}(q_i - q_j) H_{i, j}(l)
\end{equation}
was introduced by Mielke for band matrices \cite{mielk98} for $q_i=i$
and simultaneously by Knetter and 
Uhrig for many-body Hamiltonians where 
$q_i$ counts the number of elementary excitations \cite{uhrig98c, knett00a}. 
The indices $i$ and $j$ label the transition from state $j$ to state $i$. 
These states are eigenstates of an operator $Q$, which counts the number of 
elementary excitations, i.e., quasi-particles.  
The eigen value of $Q$ for the state $i$ is $q_i$.

The differences 
between the pc and the Wegner generator are discussed by means of simple 
Hamiltonians in Ref.\ \onlinecite{dusue04a}. In particular the convergence 
behavior is examined. The Wegner generator $\eta_\text{Wegner}
=[H_0,H]$ \cite{wegne94} always leads to a fixed point, 
but it does not necessarily aim at a quasi-particle picture because
degeneracies may hinder the diagonalisation. 
The pc generator is not 
sensitive to degeneracies, but the induced transformation does not always 
converge for infinite systems.

Due to the sign function the generator (\ref{pc generator}) contains only 
terms of 
the Hamiltonian that change the particle number. Terms which decrease the 
particle number acquire an additional minus sign. The resulting transformation 
yields an effective Hamiltonian for $l\rightarrow\infty$ which conserves the 
particle number (see Fig.\ \ref{pc CUT blocks}). The asymptotic behavior
\cite{mielk98,fisch10a} of 
the non-particle-conserving terms, i.e., $q_i\neq q_j$, is dominated by
\begin{equation}
\label{flow equation pc asymptotic}
\frac{\partial H_{i, j}(l)}{\partial l} \approx -
\mathrm{sign}(q_i - q_j)(H_{i, i}(l) - H_{j, j}(l)) H_{i, j}(l)\,.
\end{equation}
Because $H_{i, j}(l)$ tends to zero for $l\rightarrow\infty$ if $q_i\neq q_j$, 
the transformation tries to sort the eigenenergies $H_{i, i}$ according to the 
particle number $q_i$, i.e.,
\begin{equation}
\label{sorting of the eigen values}
\mathrm{sign}(q_i - q_j)(H_{i, i}(l) - H_{j, j}(l)) > 0
\end{equation}
holds true for large $l$.

Overlapping energies of states with different quasiparticles can hinder the 
convergence of the flow. This problem can be solved by excluding terms from the
 generator \cite{fisch10a}. The ground state or gs generator 
\begin{equation}
\label{gs generator}
\eta_{\mathrm{gs}, i, j}(l) = (\delta_{j,0} - \delta_{i,0}) H_{i, j}(l)
\end{equation}
introduced in 
Ref.\ \onlinecite{fisch10a} includes only terms that couple to the 
vacuum $|i=0\rangle$ of excitations  so that in the effective Hamiltonian 
only this vacuum  is decoupled from the remaining Hilbert space. Then it
has become the ground state. This is sufficient for the calculation of the 
hole dispersion for the spin ladder if the ground state can be
continuously linked to the state without magnetic excitations for a given hole
configuration. This is the case for a single hole at a given momentum close to
the minimum of its band.

\subsection{Real Space Restriction of the Generator}
\label{chap:gen_rest}

\begin{figure}
\begin{center}
\includegraphics[scale=0.3]{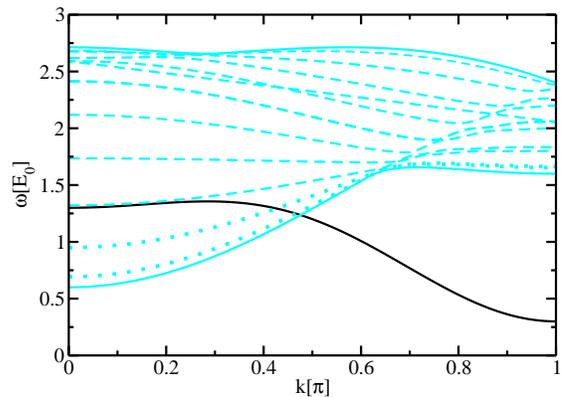}
\caption{(Color online) Schematic example for an overlap between one- and 
two-particle energies. The black line is the one-particle dispersion, the solid
 cyan (gray) lines are the boundaries of the two-particle continuum and the 
dashed lines are the energies of the two-particle states which are actually 
affected by the terms $a^{\dagger}_n a^{\dagger}_{n+\Delta n} a_m$ and 
$a^{\dagger}_m a_{n+\Delta n} a_n$. The cyan (gray) dashed lines do not pose a 
problem because they do not cross the one-particle dispersion. Convergence 
problems can only be induced by the terms that act on the states with the cyan 
(gray) dotted dispersions, which cross the one-particle dispersion.}
\label{generator truncation}
\end{center}
\end{figure}
But even the gs generator encounters convergence problems if the interaction 
between the rungs becomes too strong, i.e.,
 $x=1$, $\lambda_{\perp}=\lambda_{\parallel}>2$. Hence we have developed a 
generator adaption with a real space restriction of the generator terms.

Many terms that remain after the real space truncation of the Hamiltonian act 
on states with discrete energies within the continua (see Fig.\ 
\ref{generator truncation}). The divergence of the SCUT in case of an overlap 
is not induced by all terms that mediate between overlapping continua, but 
only by a part of these terms. Thus a stricter real space truncation of the 
Hamiltonian can induce convergence. However, a stricter truncation causes a 
larger error for the results of the SCUT.

Therefore we do not apply a stricter truncation to the Hamiltonian but we 
choose a stricter generator. We emphasize that a restriction of the generator 
does not imply an approximation. It only changes the direction of the unitary 
rotation. This restriction of the generator is not based on the total extension
 of the terms, but on the extension of the creation operators and on the 
extension of the annihilation operators separately.

Let us consider the example of an overlap between one- and two-particle 
energies shown in Fig.\ \ref{generator truncation}.  The terms 
$a^{\dagger}_n a^{\dagger}_{n+\Delta n} a^{\phantom{\dagger}}_m$ and 
$a^{\dagger}_m a^{\phantom{\dagger}}_{n+\Delta n} a^{\phantom{\dagger}}_n$ 
are responsible for transitions between the one- and the two-particle subspace.
 We choose to restrict the generator based on $|\Delta n|$, which is the 
distance between the two particles which are created or annihilated. Because 
the one-particle state does not have an extension, it is not relevant for the 
restriction of the generator. All terms that have a larger $|\Delta n|$ than a 
certain $\Delta n_{\mathrm{max}}$ are excluded from the generator. Note that 
they are still part of the Hamiltonian as long as they meet the truncation 
criteria for the Hamiltonian.

The transformation induced by this restricted generator does not try to sort 
all eigenenergies (cf.\ Eq.\ \ref{sorting of the eigen values}), but only those 
which are captured by the terms in the generator. Therefore the flow may also 
converge in case of overlapping energies. The price to be paid is that the 
subspaces affected by the omitted terms are not completely decoupled from the 
remaining Hilbert space. Thus either an additional diagonalization or
a diagrammatic analysis has to be 
applied to the effective Hamiltonian from this SCUT or the results have to be 
considered as upper limit for the actual results. The restriction of the 
generator can be applied to the pc generator or to any of its adaptions, 
e.g., to the gs generator.

\subsection{Truncation Scheme}
\label{chap:trunc_scheme}

The finite energy gap $\Delta$ of the triplons \cite{dagot88, dagot92b} is 
equivalent to a correlation function which is exponentially 
decreasing with respect to the distance. Hence a truncation in real space is 
appropriate. The extension in real space shall be used as a measure for the 
physical importance of a term of the Hamiltonian. In our quasi one-dimensional 
spin ladder the extension of a term is defined as the difference between the 
smallest and the largest rung index of the local operators within the term.

The simplest way of truncating would be one maximal extension in real space for
 all terms. Terms exceeding this limit would be omitted. But this approach is 
not reasonable in our case. The number of possible terms increases much more 
strongly with the maximal extension for terms consisting of more local 
operators. However, terms consisting of less operators are usually more 
important. Hence higher extensions should be taken into account for terms with 
less operators. For instance, the coefficient of a one-particle hopping term 
consisting of two local operators is usually larger than the coefficient of a 
two-particle interaction term consisting of four local operators if they both 
have the same extension. Therefore, for the pure triplon terms different 
maximal extensions $d_n$ are defined in units of the rung distance where $n$ 
is the total number of local operators of the term under study. Because only 
terms changing the particle number by two or conserving the particle number 
occur  in the initial magnetic Hamiltonian, only the $d_n$ 
with even $n$ are meaningful \footnote{The conservation of the parity with 
respect to the symmetry axis of the ladder forbids the change of the triplon 
number by an odd value as the parity of one triplon is odd.}.

By $h_{n'}$ we denote the maximal extension for $n'$ hole state operators. It 
does not matter if the term concerned contains additional triplon operators. 
For terms with triplon and hole state operators, $h_{n'}$ affects only the hole
 state operators. The total maximal extension $t_{n''}$ for these mixed terms 
depends on the number of triplon operators $n''$. The parameter $N_t$ denotes 
the maximal number of triplons interacting with holes. Note that $t_{n''}$ 
with odd $n''$ have to be taken into account because the triplon number is 
changed by an odd number if the hole state parity is altered.

Additionally a maximal triplon number $N$ and a maximal hole number $N_h$ are 
defined for the operators, i.e., terms affecting higher triplon numbers than 
$N$ or higher hole numbers than $N_h$ are completely omitted. This truncation 
scheme is not compulsory. Other schemes could be implemented which use 
different classifications for the groups of terms sharing the same maximal 
extension.

Since the size of the system of differential equations for the doped ladder 
grows exponentially with increasing extensions, 
the truncation for our calculations is fairly strict.
The parameters $N=4$, $d_2=10$, $d_4=6$, $d_6=4$ and 
$d_8=3$ were used for the pure triplon terms because these maximal extensions 
are sufficient for the undoped case up to $x=1$.

For the terms including hole operators the truncation used for the SCUT is 
given by $N_h=1$, $N_t=2$, $h_2=3$, $t_1=6$, $t_2=6$, $t_3=5$, $t_4=5$, 
$t_5=4$ and $t_6=4$. All results shown in the sequel are obtained with these 
parameters unless stated otherwise.

\section{Hole Dispersions without Ring Exchange}
\label{chap:hole disp without ring exc}

\subsection{Calculations with Unrestricted Generators}
\label{chap:calc_with_unr_gen}

\begin{figure}
\begin{center}
\includegraphics[scale=0.3]{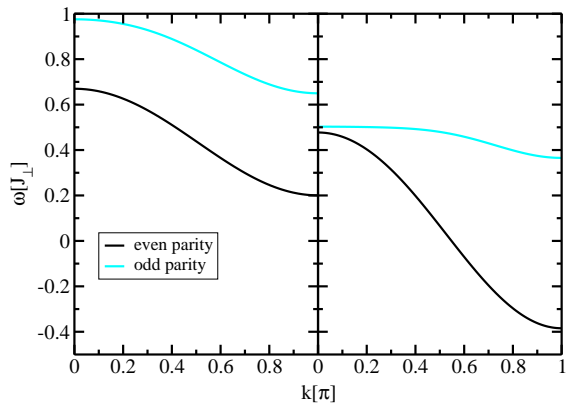}
\caption{(Color online) One-hole dispersion for $x=x_{\square}=0$ calculated 
with the pc generator; left: $\lambda=0.25$, right: $\lambda=0.5$. The curves 
coincide well with the series expansion results from \onlinecite{oitma99}.}
\label{x=0 lambda=0.25,0.5}
\end{center}
\end{figure}
The dispersion of a single hole in the absence of triplons can be easily 
derived from the effective Hamiltonian by means of Fourier transformation. 
The part of the Hamiltonian to be diagonalised
\begin{equation}
H_{\mathrm{1h}} = \sum_{d=-h_2}^{h_2} \sum_{\tau, \sigma, n} b_d \, 
a^{\dagger}_{\tau, \sigma, n+d} a^{\phantom{\dagger}}_{\tau, \sigma, n}
\end{equation}
contains the one-hole terms restricted by $h_2$ and characterised by the 
coefficients $b_d$. The one-hole dispersion
\begin{equation}
\label{fourier series}
\omega_{\mathrm{1h}, \tau}(k) = b_0 + \sum_{d=1}^{h_2} 2 b_d \cos(d k)
\end{equation}
only depends on the parity $\tau$ and it is degenerate concerning the spin 
$\sigma$.

If $\lambda_{\parallel}$ and $\lambda_{\perp}$ are small while
 $x=x_{\square}=0$, the dispersion reads
\begin{equation}
\label{hole dispersion for small parameters}
\frac{\omega_{\mathrm{1h}, \pm1}(k)}{J_{\perp}} = \frac{3}{4} \mp 
\lambda_{\perp} + \lambda_{\parallel}\cos(k) + \mathcal{O}(\lambda_{\perp}^2, 
\lambda_{\parallel}^2, \lambda_{\perp} \lambda_{\parallel}).
\end{equation}
In the following we only consider the isotropic case $\lambda_{\perp} = 
\lambda_{\parallel} = \lambda$. For small values of the parameters $x$, 
$x_{\square}$ and $\lambda$ the deviations from Eq.\ 
(\ref{hole dispersion for small parameters}) are actually small. In Fig.\ 
\ref{x=0 lambda=0.25,0.5} two values of $\lambda$ are considered for 
$x=x_{\square}=0$. The result for $\lambda=0.25$ already exhibits deviations 
from the relation (\ref{hole dispersion for small parameters}) for small 
hopping constants. Both bandwidths are smaller than $2\lambda=0.5$. The odd 
band is narrower than the even band. Moreover the odd band is shifted upwards 
by less than $\lambda$, while the even band is shifted downwards by slightly 
more than $\lambda$.

\begin{figure}
\begin{center}
\includegraphics[scale=0.3]{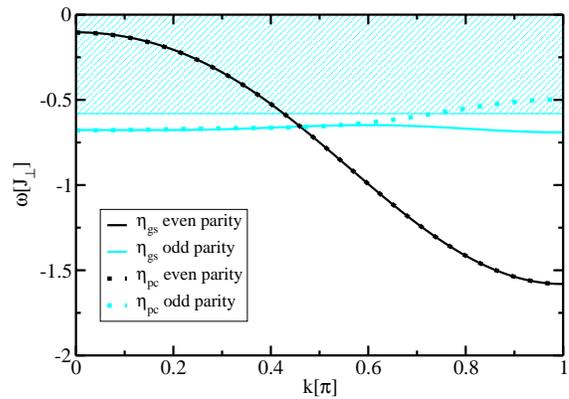}
\caption{(Color online) One-hole dispersion for $x=x_{\square}=0$, $\lambda=1$ 
calculated with the gs generator and the pc generator. The results for the 
even bands coincide so that the curves cannot be distinguished. The cyan (gray)
 shaded area is the continuum formed by one triplon and one even hole state.}
\label{x=0 lambda=1}
\end{center}
\end{figure}

The increase of $\lambda$ to $0.5$ yields a result with obvious deviations 
relative to Eq.\ 
(\ref{hole dispersion for small parameters}). The odd band becomes 
lower and narrower with growing $\lambda$ in this region. The cosine shape of 
both dispersions is -- for $\lambda=0.25$ as well as for $\lambda=0.5$ -- not 
deformed by higher harmonics. The results of the series expansion 
\cite{oitma99} exhibit the same behavior for these parameters in good 
agreement with the SCUT results. However, it has to be pointed out that the 
convergence of the SCUT is much worse for $\lambda=0.5$ than for 
$\lambda=0.25$. The residual off-diagonality (ROD) defined in 
Sect.\ \ref{chap:SCUT} and used as a measure for the convergence is decreasing 
very slowly for $\lambda=0.5$. While the ROD is smaller than $10^{-6}$ at 
$l J_{\perp} = 200$ for $\lambda=0.25$, it is still larger than $10^{-3}$ at 
$l J_{\perp} = 200$ for $\lambda=0.5$. Both RODs are decreasing exponentially 
for large $l$. Theses RODs are not shown here.

For $x=x_{\square}=0$, $\lambda=1$ the pc generator yields results for the even
 band that still agree very well with the series expansion. For the odd band 
there are no series expansion results available. Fig.\ \ref{x=0 lambda=1} shows
 the comparison of the results for the pc generator and for the gs generator. 
The even band results lie on top of each other, but the odd band exhibits 
deviations for $k\gtrsim0.57\pi$. While the gs generator produces an almost 
featureless dispersion (compared to the even band), the pc generator causes a 
pronounced maximum at $k=\pi$. These differences are due to the position of the
 lower boundary of the continuum formed by one triplon and one even hole 
state\footnote{The boundaries of this continuum are constant because the 
triplon dispersion is also constant for $x=x_{\square}=0$.} (see also Fig.\ 
\ref{x=0 lambda=1}). This is the odd continuum due to the odd parity of the 
triplon. An overlap between the odd dispersion and this continuum is present 
for the pc result. The deviating odd band from the gs generator avoids this 
overlap. For both generators the even and the odd band cross at 
$k\approx0.46\pi$. While the even band keeps its cosine shape, the second 
harmonic for the odd band is no longer negligible for both the pc and the 
gs generator.

\begin{figure}
\begin{center}
\includegraphics[scale=0.3]{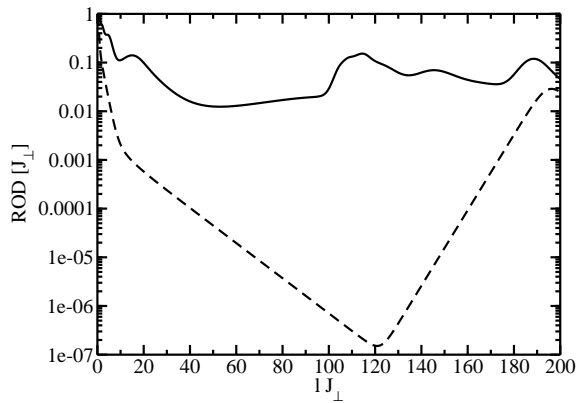}
\caption{ROD for the SCUT for $x=x_{\square}=0$, $\lambda=1$ induced by the pc 
generator (solid line) and by the gs generator (dashed line).}
\label{x=0 lambda=1 ROD}
\end{center}
\end{figure}
For $x=x_{\square}=0$, $\lambda=1$ the transformation does not converge for the
 pc generator, while the gs generator induces convergence (see 
Fig.\ \ref{x=0 lambda=1 ROD}). The overlap hinders the convergence for the pc 
generator. The kink at $lJ_{\perp}\approx120$ in the gs ROD is probably due to 
numerical inaccuracies that are amplified via a feedback within the flow 
equation. Since the relative ROD is already $\approx10^{-7}$ 
at the kink we can stop 
the transformation at this point and neglect the remaining off-diagonal terms, 
i.e., we consider the transformation to be converged. 
For the calculation of the hole
 dispersions the pc SCUT was stopped at $lJ_{\perp}\approx50$ where the ROD is 
still $\approx10^{-2}$ and thus 
not negligible. Therefore the result from the gs 
generator is more trustable than the result from the pc generator. The 
parameters $x=x_{\square}=0$, $\lambda=2$ lead to divergence for either 
generator.

\subsubsection{Finite Coupling Ratio $x=J_\parallel/J_\perp$}

Here we want to consider $x>0$. For $\lambda=0$ the energy of the hole states, 
which is independent of momentum $k$ and parity $\tau$, decreases with 
increasing $x$. Already for small finite $\lambda$ the deviations 
from the simple cosine shape appear  for the odd band. This can be seen in 
Fig.\ \ref{x=0.5 lambda=0.25,0.5} where the one-hole dispersions for $x=0.5$ 
and $\lambda=0.25$ are depicted. The odd band agrees well with the series 
expansion; only slight deviations at $k=0$ occur. For the even band no series 
expansion data are available.

\begin{figure}
\begin{center}
\includegraphics[scale=0.3]{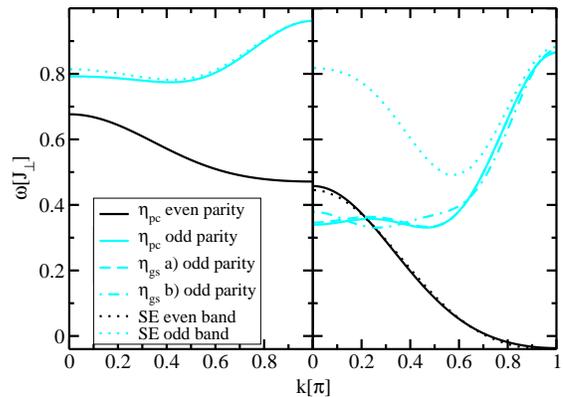}
\caption{(Color online) Left panel: One-hole dispersion for $x=0.5$, 
$x_{\square}=0$, $\lambda=0.25$ calculated with the pc generator. The series 
expansion result for the odd band is also shown, while the result for the even 
band is not available. Right panel: One-hole dispersion for $x=0.5$, 
$x_{\square}=0$, $\lambda=0.5$ calculated with the pc generator and the gs 
generator. For case a) in the legend
 the usual truncation was used. For case b) in the legend
$h_2$ was increased to 
$4$, $t_1$ and $t_2$ to $8$ as well as $t_3$ and $t_4$ to $6$. The dotted 
curves are extrapolations from the series expansion data.}
\label{x=0.5 lambda=0.25,0.5}
\end{center}
\end{figure}

For $x=0.5$ and $\lambda=0.5$ , see also Fig.\ \ref{x=0.5 lambda=0.25,0.5}, 
the result for the even band agrees again well with the series expansion result
 \cite{oitma99}, but the odd band behaves differently. Only for large $k$ the 
behavior is similar although also in this region the SCUT result is slightly 
lower. The series expansion result exhibits a local maximum at $k=0$ and a 
global minimum at $k\approx0.58\pi$, while the SCUT result hardly changes in 
the region $0<k<\frac{\pi}{2}$. The gs generator yields the same result as the 
pc generator apart from minimal deviations ($<2\%$) at $k=0$. However, the gs 
generator allows us to extend the truncation scheme: $h_2$ was increased to 
$4$, $t_1$ and $t_2$ to $8$ as well as $t_3$ and $t_4$ to $6$. In the result 
the shape of the odd band changes mainly for small $k$. Thus the odd 
dispersion still changes with increasing maximal extensions.

\begin{figure}
\begin{center}
\includegraphics[scale=0.3]{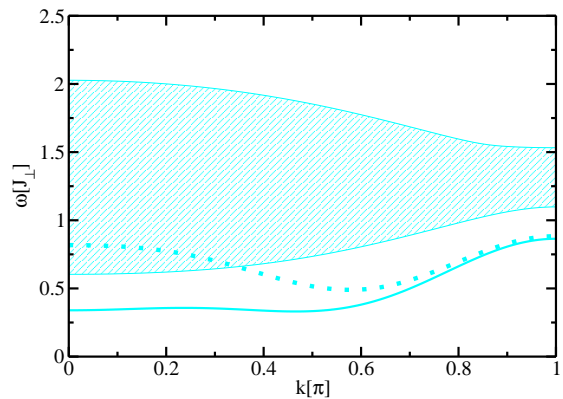}
\caption{(Color online) Comparison of the odd hole dispersion and the 
continuum formed by one triplon and one even hole state for 
$x=0.5$, $x_{\square}=0$, $\lambda=0.5$. The odd hole dispersion was derived 
from SCUT using the pc generator (cyan/gray solid) and by series expansion 
(cyan/gray dotted). The cyan (gray) shaded area is the continuum (derived from 
the same SCUT calculation).}
\label{x=0.5 lambda=0.5 continuum}
\end{center}
\end{figure}

We can understand these differences between SCUT and series expansion by 
looking at the position of the continuum formed by one triplon and one even 
hole state (see Fig.\ \ref{x=0.5 lambda=0.5 continuum}). The series expansion 
result crosses the lower boundary of this continuum at $k\approx0.34\pi$. 
Assuming that the series expansion yields a result close to the actual odd 
hole dispersion it is cogent that the SCUT based on the pc generator is not 
able to sort the eigen values properly for small $k$ because creating one 
triplon does not necessarily increase the energy. 

Since the problematic overlap
 concerns the zero- and the one-triplon space, the use of the gs generator is 
no remedy here. This may come as a surprise at first since
it was shown in Ref.\ \onlinecite{fisch10a} that the gs generator always 
implies a robust flow. This statement applies if the hole is
put in its band minimum because this is  by construction the lowest 
eigen state of the ladder doped with one hole. But for other energies
of the hole dispersion this does not need to be the case so that
the energy of one hole may lie higher than the energy of one hole
and one triplon.
Hence the gs generator yields mainly the same result and 
suffers from the same problems as the pc generator.
A robust extrapolation of the series can still yield reasonable results
because the series is not fully sensitive to the overlap of states
at higher values of the expansion parameters.

\begin{figure}
\begin{center}
\includegraphics[scale=0.3]{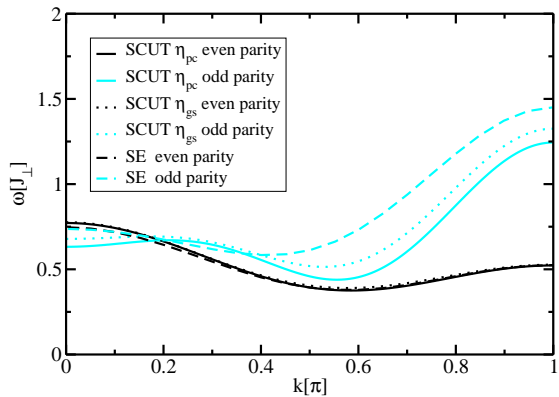}
\caption{(Color online) One-hole dispersion for $x=1$, $x_{\square}=0$, 
$\lambda=0.5$ calculated with the pc generator and the gs generator. The 
dashed curves are extrapolations from the series expansion data.}
\label{x=1 lambda=0.5}
\end{center}
\end{figure}
\begin{figure}
\begin{center}
\includegraphics[scale=0.3]{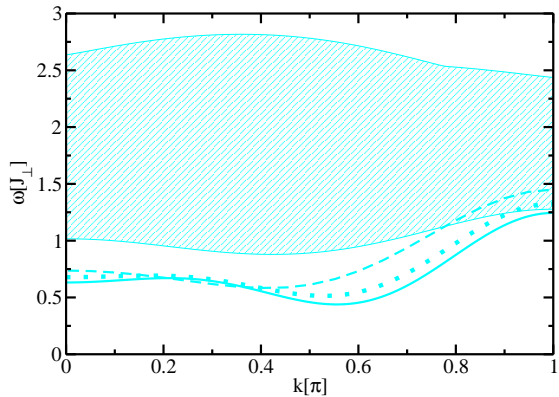}
\caption{(Color online) Comparison of the odd hole dispersion and the 
continuum formed by one triplon and one even hole state for $x=1$, 
$x_{\square}=0$, $\lambda=0.5$. The odd hole dispersion was derived from SCUT 
using the pc generator (cyan/gray solid), using the gs generator (cyan/gray 
dotted) and by series expansion (cyan/gray dashed). The cyan (gray) shaded 
area is the continuum (derived from the same pc SCUT calculation).}
\label{x=1 lambda=0.5 continuum}
\end{center}
\end{figure}
For $x=1$ and $\lambda=0.5$ (see Fig.\ \ref{x=1 lambda=0.5}) the SCUT results 
for the odd band again show distinct deviations from the series expansion 
results. The even band exhibits a maximal deviation at $k=0$
of only $\approx1\%$. The deviations of the energetically higher odd band are
larger, but still well tolerable for the gs generator. 
All in all the deviations are not as 
pronounced as for $x=0.5$ and $\lambda=0.25$. A comparison with the continuum 
formed by one even hole and one triplon (see 
Fig.\ \ref{x=1 lambda=0.5 continuum}) shows that an overlap exists around 
$k=\pi$ for the series expansion result and for the gs result. This explains 
again the deviations in this region. As this overlap is not as strong as the 
overlap for $x=0.5$ and $\lambda=0.5$ the deviations are smaller. 
Even if there is no actual overlap, the continuum is at least very close. 
Hence it is to be expected that the actual odd hole dispersion is lowered for 
$k\approx\pi$ due to this reason.

\begin{figure}
\begin{center}
\includegraphics[scale=0.3]{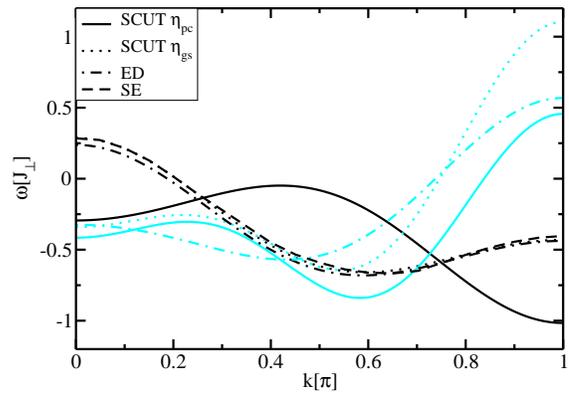}
\caption{(Color online) One-hole dispersion for $x=1$, $x_{\square}=0$, 
$\lambda=1$ calculated with the pc generator and the gs generator. The dashed 
curve is the result from the series expansion, for which only the even band is 
available. The dashed-dotted curves are exact diagonalization results.}
\label{x=1 lambda=1}
\end{center}
\end{figure}

For $x=1$ and $\lambda=1$ (see Fig.\ \ref{x=1 lambda=1}) no one-hole 
dispersion with odd parity is available from series expansion.
Thus we look at the even band including results from exact 
diagonalization. For the exact diagonalization a finite ladder with 
$14$ rungs was examined and the resulting eigenenergies were fitted by a sum
of three cosine terms for both bands, see Eq.\ (\ref{fourier series}). 
The pc result for the even band differs completely from the gs result. 
For the odd band the pc result lies below the gs result and the deviations grow
 with increasing momentum. The comparison with the series expansion and with 
the exact diagonalization suggests that the gs result is more reliable because 
the deviations are smaller than $1\%$ for the even band. Also for the odd band 
the gs result is closer to the exact diagonalization result. The odd band is 
more difficult to assess due to the vicinity to the odd continuum.

It should be noted that the diagrammatic approach from 
Ref.\ \onlinecite{jurec01} yields bands that show qualitative deviations from 
the SCUT and from the exact diagonalization concerning the shape for small $k$.
The authors state that this regime actually exceeds the applicability of
their approach. The quantum Monte Carlo result 
from Ref.\ \onlinecite{brunn01} is in very good agreement with our result. For 
the even band the agreement is even excellent and comparable to the 
agreement between SCUT and exact diagonalization.

\begin{figure}
\begin{center}
\includegraphics[scale=0.3]{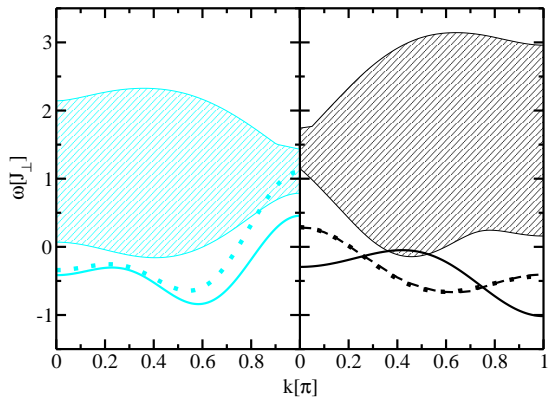}
\caption{(Color online) Comparison of the hole dispersions and the continua 
formed by one triplon and one hole state for $x=1$, $x_{\square}=0$, 
$\lambda=1$. Left: The odd hole dispersion was derived from SCUT using the pc 
generator (cyan/gray solid) and using the gs generator (cyan/gray dotted). The 
cyan (gray) shaded area depicts the continuum with one even hole and one 
triplon derived from the same gs SCUT calculation. Right: The even hole 
dispersion was derived from SCUT using the pc generator (black solid), using 
the gs generator (black dotted) and by series expansion (black dashed). The 
black shaded area depicts the continuum with one odd hole and one triplon 
derived from the same gs SCUT calculation.}
\label{x=1 lambda=1 continuum}
\end{center}
\end{figure}

The odd band result from the gs SCUT enters the continuum formed by one triplon
 and one even hole state for $k\gtrsim0.83\pi$ (see 
Fig.\ \ref{x=1 lambda=1 continuum}), while the pc result lies always below this
 continuum. This overlap is also present in the exact diagonalization result 
and in the quantum Monte Carlo result \cite{brunn01}. A comparison of the even 
one-hole dispersion and the even continuum formed by one triplon and one 
odd hole state (see also Fig.\ \ref{x=1 lambda=1 continuum}) supports the 
assumption that the gs result is more reliable than the pc result. The 
dispersion induced by the gs SCUT exhibits a shape that appears as if it were 
formed by the lower boundary of the approaching continuum. This is plausible 
because the vicinity of the continuum lowers the dispersion. The pc result, 
however, stays away from the continuum at $k=0$ and at the boundary of the 
Brillouin zone. The overlap with the continuum around 
$k\approx 0.44\pi$ occurs in the pc calculation only due to
truncation.

\begin{figure}
\begin{center}
\includegraphics[scale=0.3]{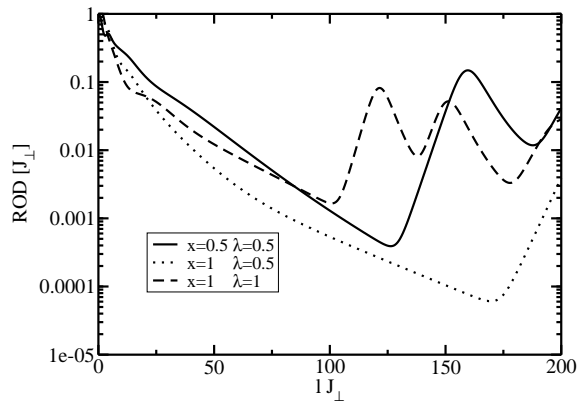}
\caption{ROD for the SCUT calculated with the pc generator for various 
parameters.}
\label{ROD comparison}
\end{center}
\end{figure}

At this point we compare the convergence behavior for $x=0.5$, $\lambda=0.5$; 
$x=1$, $\lambda=0.5$ and $x=1$, $\lambda=1$ for  the pc generator. The 
corresponding RODs are depicted in Fig.\ \ref{ROD comparison}. All three curves
 exhibit a kink with non-converging behavior afterwards. This is typical for 
cumulating rounding errors because of a symmetry breaking due to numerical 
inaccuracies. Such a symmetry breaking is not unlikely because the full spin 
symmetry could not be used explicitly.

It is surprising that the ROD for $x=1$, $\lambda=0.5$ 
reaches a lower minimum than the one for the lower
parameter set $x=0.5$, $\lambda=0.5$. This seems to contradict
the expectation that larger values of the interdimer processes
imply a more difficult CUT. But inspecting the slope of
the RODs in Fig.\ \ref{ROD comparison}  before reaching 
their minima one sees that these slopes, i.e., the convergence
velocities, are indeed largest for the smallest values for
the interdimer processes.

\subsubsection{Strong Hopping $\lambda=t/J_\perp >1$}

\begin{figure}
\begin{center}
\includegraphics[scale=0.3]{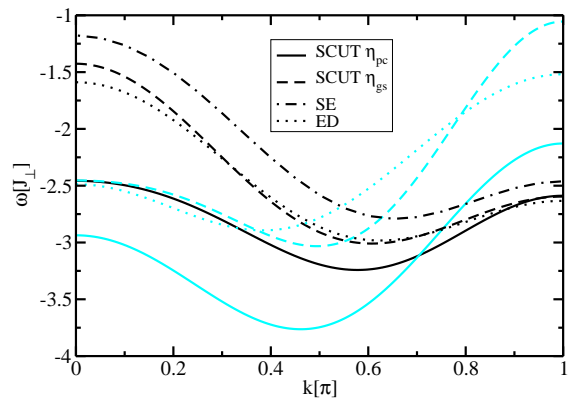}
\caption{(Color online) One-hole dispersion for $x=1$, $x_{\square}=0$, 
$\lambda=2$ calculated with the pc generator and the gs generator. The graph 
also shows the exact diagonalization data (ED) and the series expansion data 
(SE) \cite{oitma99}. For the latter only the even band is available. The even 
band results are black, the odd band results are cyan (gray).}
\label{x=1 lambda=2}
\end{center}
\end{figure}

For $x=1$ and $\lambda>1$ the parameters are entering a region which is 
expected to reflect realistic relations of the constants in the telephone 
number compounds. The results for $x=1$ and $\lambda=2$ are shown in 
Fig.\ \ref{x=1 lambda=2}. Again we compare with results from series 
expansion and from  exact diagonalization for $14$ rungs.

Approximate analytic results were obtained in Ref.\ \onlinecite{sushk99} by 
perturbation theory improved by a variational ansatz. These results lie even 
above the series expansion results but confirm the qualitative shape for the 
even band. The pc result for both the even and the odd band is again very 
distinct from the other results comparable to $x=1$, $\lambda=1$. The gs 
result, however, is in better agreement with the data from the series 
expansion and especially with the exact diagonalization result in accordance 
with our previous observations.

Apart from the pc result the dispersions exhibit the same features. The even 
band has a global maximum at $k=0$ and a local maximum at $k=\pi$, while it is 
vice versa for the odd band. Because both bands lie in the same energy range, 
they cross in the middle between $k=0$ and $k=\pi$. The exact diagonalization 
predicts the crossing to be at $k\approx0.48\pi$, but the gs SCUT finds the 
crossing at $k\approx0.55\pi$. The even band calculated by series expansion is 
located above both the series expansion and the gs SCUT result for all $k$. 
This is further evidence that the extrapolation used to correct the bare 
series underestimates the lowering of the band induced by the hybridisation 
with the hole-triplon continuum.

\begin{figure}
\begin{center}
\includegraphics[scale=0.3]{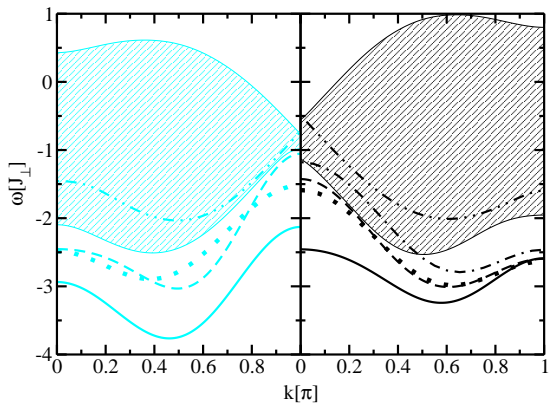}
\caption{(Color online) Comparison of the hole dispersions and the continua 
formed by one triplon and one hole state for $x=1$, $x_{\square}=0$, 
$\lambda=2$. Left: The odd hole dispersion was derived from pc SCUT (cyan/gray 
solid), gs SCUT (cyan/gray dashed) and by exact diagonalization (cyan/gray 
dotted). The cyan (gray) shaded area depicts the continuum with one even hole 
and one triplon (derived from the same gs SCUT calculation). The cyan (gray) 
dash dot dot line is the lower boundary of the continuum formed by two triplons
 and one odd hole. Right: The even hole dispersion was derived from pc SCUT 
(black solid), gs SCUT (black dashed), by exact diagonalization (black dotted) 
and by series expansion (black dashed-dotted). The black shaded area depicts 
the continuum with one odd hole and one triplon (derived from the same gs 
SCUT calculation). The black dash dot dot line is the lower boundary of the 
continuum formed by two triplons and one even hole.}
\label{x=1 lambda=2 continuum}
\end{center}
\end{figure}
Let us consider the continua formed by one hole and one triplon. The continua 
consisting of one hole and one triplon are compared to the one-hole 
dispersions in Fig. \ref{x=1 lambda=2 continuum}. The continua do not overlap 
with the hole dispersions, but they are very close to them. The only 
exception is the series expansion result for the odd band which 
actually exhibits an 
overlap with the continuum formed by a triplon and an even hole state.

\begin{figure}
\begin{center}
\includegraphics[scale=0.3]{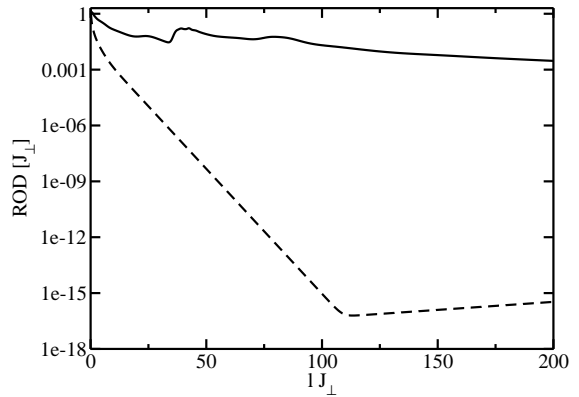}
\caption{ROD for the SCUT for $x=1$, $x_{\square}=0$, $\lambda=2$ induced by 
the pc generator (solid line) and by the gs generator (dashed line).}
\label{x=1 lambda=2 ROD}
\end{center}
\end{figure}
The behavior of the ROD yields further evidence why the gs results should be 
preferred to the pc results for this parameter regime. 
Fig.\ \ref{x=1 lambda=2 ROD} depicts the evolution of the ROD during the flow 
for both generators. Not only that the pc SCUT converges extremely slowly, the 
shape of the curve for $l J_{\perp}<100$ is an indicator for a problem with 
respect to the sorting of the eigenenergies. It is a generic behavior of the 
ROD that the sorting of the eigen values (cf.\ 
Eq.\ (\ref{sorting of the eigen values})) is reflected by features for small 
values of $l$. When the sorting is completed the ROD decreases exponentially 
with a constant rate. For the gs generator the decrease of the 
ROD attains this rate not later than at $l J_{\perp}=5$. Before this point the 
decrease is slower \footnote{Even if the ROD exibits a small hump at small 
values of $l$ before it decreases with a constant rate, the sorting usually 
does not pose a problem.}. The kink of the gs ROD at $l J_{\perp} \approx 110$ 
with the subsequent upturn is again most probably due to cumulating numerical 
inaccuracies. This is no real problem because the gs ROD has already fallen 
below a value of less than $10^{-16}$ at $l J_{\perp} \approx 110$ and can 
hence be neglected. However, the pc ROD exhibits several humps and a 
pronounced rise at $l J_{\perp}=34$ before a decrease with a constant rate is 
achieved. This is a typical indication for a suppressed divergence that would 
actually occur for a less strict truncation. If such a behavior occurs
the transformation is susceptible to truncation errors. 
The physical origin of these problems is 
 the strong overlap between the one-hole-one-triplon 
continuua and the one-hole-two-triplon continua (see Fig.\ 
\ref{x=1 lambda=2 continuum}).

\subsection{Calculations with Restricted Generators}
\label{chap:calc_with_restr_gen}

Results from
 exact diagonalization are available for $x=1$ and $\lambda=3$, but even 
the gs generator does not induce convergence for this case. However, if we 
apply the generator restriction introduced in Sect.\ \ref{chap:gen_rest} to 
the gs generator, it yields convergence for 
$\Delta n_{\mathrm{max}}\leqslant2$. 
Then some coupling between one hole and one hole plus one triplon 
$1\text{h}\to 1\text{h}+1\text{t}$
is not
eliminated completely. Because the Hamiltonian is not 
diagonalised with respect to the terms  omitted from the generator, 
the hole dispersions we obtain from a Fourier transformation are only upper 
limits for the actual result. But this is a minor problem 
since the remaining matrix elements are small. Another way of improvement
(not followed here) would be to use self-consistent Born approximation to deal
with the remaining coupling $1\text{h}\to 1\text{h}+1\text{t}$.

\begin{figure}
\begin{center}
\includegraphics[scale=0.3]{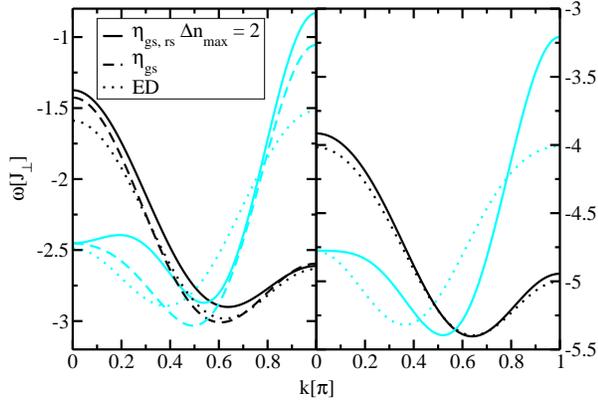}
\caption{(Color online) Comparison of the one-hole dispersions from the 
restricted generator with the gs and the exact diagonalization results. 
Left panel: $x=1$, $x_{\square}=0$, $\lambda=2$. Right panel: 
$x=1$, $x_{\square}=0$, $\lambda=3$. The black lines represent the even hole 
dispersion and the cyan (gray) lines represent the odd hole dispersion.}
\label{truncated generator results}
\end{center}
\end{figure}

 Here we compare the results for upper limits 
from the restricted generator for $x=1$ and $\lambda=2$ with the gs results 
before we investigate the results for $x=1$ and $\lambda=3$. The left panel of 
Fig.\ \ref{truncated generator results} shows the comparison of the one-hole 
dispersions for $x=1$ and $\lambda=2$. We see that for 
$\Delta n_{\mathrm{max}}\leqslant2$ the upper boundary from the restricted gs 
generator is close to the result from the full gs generator. In the right 
panel of Fig.\ \ref{truncated generator results} the results from the 
restricted gs generator are compared to the exact diagonalization results for 
$x=1$ and $\lambda=3$. For the even hole dispersion the agreement between the 
result from the restricted gs generator and the exact diagonalization result 
is almost perfect. Also the agreement for the odd hole dispersion is good. 
\footnote{Considering the momenta where the triplon 
dispersion and the triplon-hole-continuum can be clearly distinguished, the 
deviations between SCUT and exact diagonalisation are diminishing even further
with growing  system size, see the finite size scaling in Appendix 
\ref{app:fin_size_scal}.}
The deviations are comparable to the deviations of the result by the full gs 
generator from the exact diagonalization result for $x=1$ and $\lambda=2$.

\begin{figure}
\begin{center}
\includegraphics[scale=0.3]{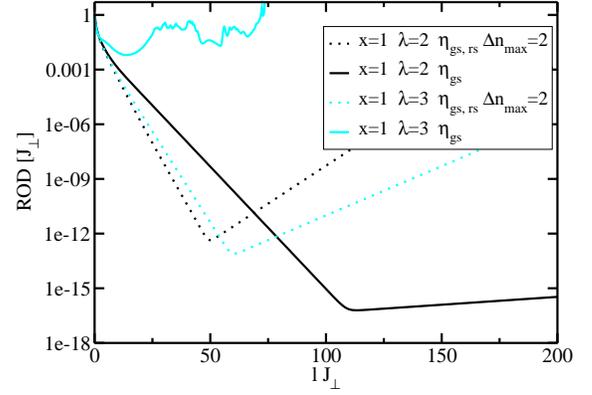}
\caption{(Color online) ROD for the SCUT calculated with the full gs generator 
and with the restricted gs generator for different parameters.}
\label{truncated generator rod}
\end{center}
\end{figure}
The investigation of the ROD shows that the restricted gs generator yields a 
faster convergence than the full gs generator for $x=1$ and $\lambda=2$ 
(see Fig.\ \ref{truncated generator rod}). For $x=1$ and $\lambda=3$ the ROD 
diverges for the full generator, while the restricted generator induces 
convergence (see also Fig.\ \ref{truncated generator rod}). Note that the 
kinks of the RODs in Fig.\ \ref{truncated generator rod} with an increase 
afterwards are again probably due to a feedback of numerical inaccuracies. 
But all the kinks appear at values where the ROD is already smaller than 
$10^{-12}$. So the flow can be considered to be 
converged at the kinks for practical purposes.

The terms left out from the restricted generator still contribute 
to the Hamiltonian after the transformation. These contributions yield an 
estimate of the difference between the actual energy and the upper boundary 
for the energy resulting from the restricted generator. For the $432$ terms of 
the form $a^{\dagger}_{\tau, \sigma, n} t^{\dagger}_{\alpha, n+\Delta n} 
a^{\phantom{\dagger}}_{\tau', \sigma', m}$ or $a^{\dagger}_{\tau, \sigma, m} 
t^{\phantom{\dagger}}_{\alpha, n} a
^{\phantom{\dagger}}_{\tau', \sigma', n+\Delta n}$ the sum over their squared 
coefficients is $\approx0.48\,\mathrm{J}_{\perp}^2$; its square root 
yields $\approx0.16\,\mathrm{J}_{\perp}$.

\begin{figure}
\begin{center}
\includegraphics[scale=0.3]{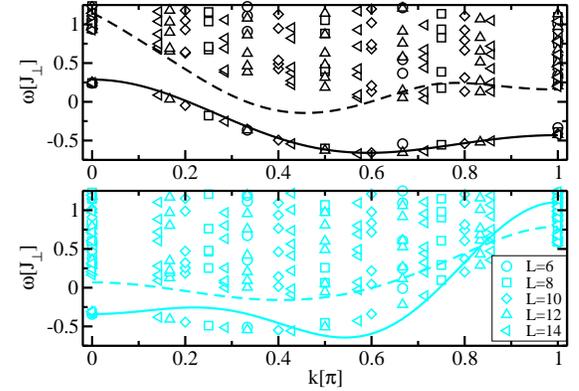}
\caption{(Color online) Comparison of gs SCUT (solid lines) and exact 
diagonalization for various finite ladders with $L$ rungs (discrete points) for
 $x=1$, $\lambda=1$. The lower boundaries of the continua calculated by gs 
SCUT are shown as dashed lines. The results for the even hole state are black, 
the results for the odd hole state are cyan (gray).}
\label{ED AL x=1 lambda=1}
\end{center}
\end{figure}
To understand the deviations between SCUT and exact diagonalization we also 
compare our results to the complete spectrum from exact diagonalization. 
In Fig.\ \ref{ED AL x=1 lambda=1} the results of the exact diagonalization for 
various finite ladders with $L$ rungs are compared to the results from the gs 
generator for $x=1$ and $\lambda=1$. For the even band the 
lowest lying eigen value can be clearly distinguished from the larger 
eigen values, which are the precursor of the continuum. This holds true for all 
momenta. In the case of the odd band the lowest eigen value is very close to 
the higher ones for large momenta. Our calculation of the continuum predicts 
that the dispersion merges with the continuum in this region. Also the quantum 
Monte Carlo result for the spectral weight \cite{brunn01} exhibits no peak 
below the continuum around $k\approx\pi$.

Our result for the even band is in excellent agreement with the exact 
diagonalization result. Around $k\approx0.25\pi$ our result for the odd band 
lies above the exact diagonalization result while it lies below the exact 
diagonalization result around $k\approx0.6\pi$.
\begin{figure}
\begin{center}
\includegraphics[scale=0.3]{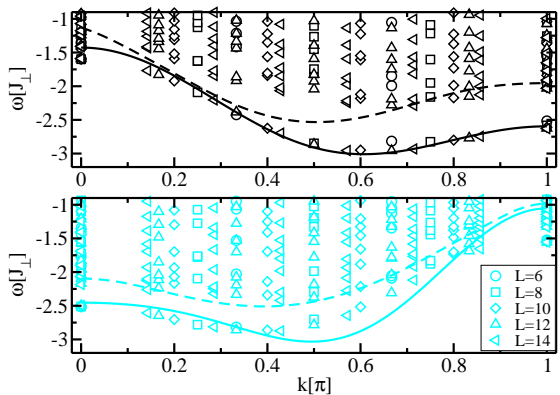}
\caption{(Color online) Comparison of gs SCUT (solid lines) and exact 
diagonalization for various finite ladders with $L$ rungs (discrete points) 
for $x=1$, $\lambda=2$. The lower boundaries of the continua calculated by gs 
SCUT are shown as dashed lines. The results for the even hole state are black, 
the results for the odd hole state are cyan (gray).}
\label{ED AL x=1 lambda=2}
\end{center}
\end{figure}
In Fig.\ \ref{ED AL x=1 lambda=2} we compare the results of the exact 
diagonalization with the results from the gs generator for $x=1$ and 
$\lambda=2$. For the even band for large momenta the lowest 
lying eigen value is clearly distinguishable from the larger eigen values, 
which are again the precursor of the continuum. The same is true for the 
odd band for small momenta. But for small momenta in case of the even band and 
for large momenta in case of the odd band the lowest eigen value is very close 
to the higher ones. Hence the distinction between hole dispersion and continuum
 becomes difficult in these regions. It is even questionable if they are 
actually distinct or if the dispersion merges with the continuum. In 
these regions the deviations between SCUT and exact diagonalization are the 
largest. We conclude that the SCUT suffers from truncation errors if the 
dispersion runs close to continua or even enters them. Also the lower boundary 
of the continuum that we calculated from the gs result is higher at $k=0$ for 
the even band and at $k=\pi$ for the odd band than we would expect from the 
exact diagonalization data.

\begin{figure}
\begin{center}
\includegraphics[scale=0.3]{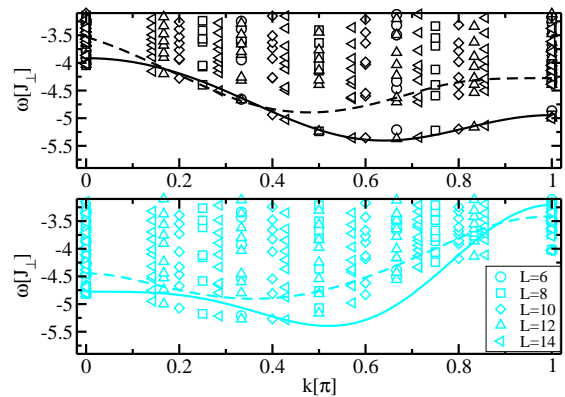}
\caption{(Color online) Comparison of SCUT from the restricted gs generator 
with $\Delta n_{\mathrm{max}}=2$ (solid lines) and exact diagonalization for 
various numbers or rungs $L$ (symbols) for $x=1$, $\lambda=3$.
 The dashed lines are estimates for the lower boundaries of the continua based 
on the result from the restricted gs generator. The results for the even hole 
state are black, the results for the odd hole state are cyan (gray).}
\label{ED AL x=1 lambda=3}
\end{center}
\end{figure}
For $x=1$ and $\lambda=3$ the exact diagonalization results are shown in 
Fig.\ \ref{ED AL x=1 lambda=3} for various finite ladders with $L$ rungs. The 
odd hole dispersion is more difficult to distinguish from the continuum than 
for $x=1$ and $\lambda=2$. Also for small momenta the lowest eigen value is 
very close to the higher ones. The hole dispersions 
from the restricted gs generator with $\Delta n_{\mathrm{max}}=2$ is also 
depicted in Fig.\ \ref{ED AL x=1 lambda=3}. In the region 
$0.5\pi\lesssim k\lesssim0.8\pi$ this dispersion lies below the exact 
diagonalization data.

\section{Hole Dispersions with Ring Exchange}
\label{chap:hole disp with ring exc}

\begin{figure}
\begin{center}
\includegraphics[scale=0.3]{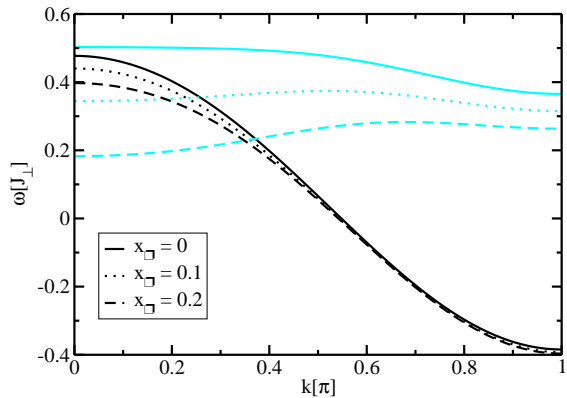}
\caption{(Color online) One-hole dispersions with even parity (black) and odd 
parity (cyan/gray) for $x=0$, $\lambda_{\perp} = \lambda_{\parallel} = 0.5$
 and various values for $x_{\square}$.}
\label{x=0 lambda=0.5 ring exchange}
\end{center}
\end{figure}

\begin{figure}
\begin{center}
\includegraphics[scale=0.3]{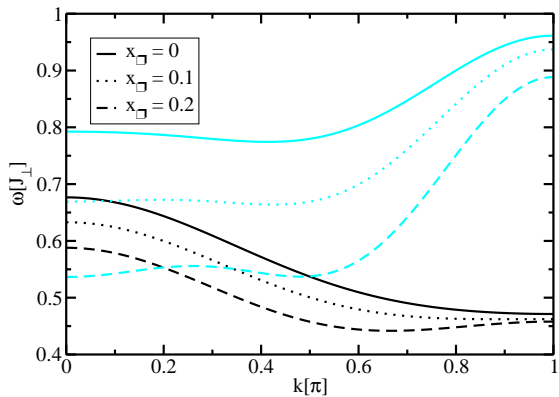}
\caption{(Color online) One-hole dispersions with even parity (black) and odd 
parity (cyan/gray) for $x=0.5$, $\lambda_{\perp} = \lambda_{\parallel} = 0.25$ 
and various values for $x_{\square}$.}
\label{x=0.5 lambda=0.25 ring exchange}
\end{center}
\end{figure}
\begin{figure}
\begin{center}
\includegraphics[scale=0.3]{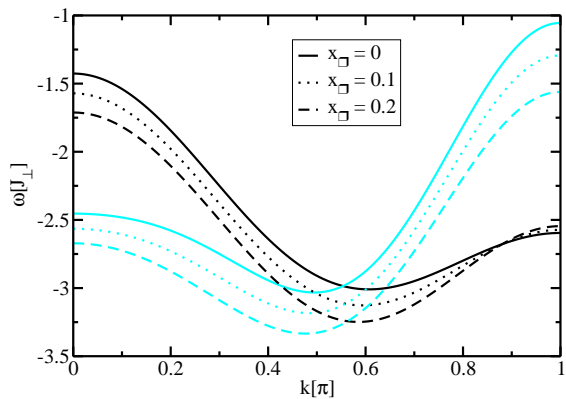}
\caption{(Color online) One-hole dispersions with even parity (black) and odd 
parity (cyan/gray) for $x=1$, $\lambda_{\perp} = \lambda_{\parallel} = 2$ and 
various values for $x_{\square}$.}
\label{x=1 lambda=2 ring exchange}
\end{center}
\end{figure}
Since the ring exchange is needed for an adequate description of the 
experimentally available systems, we also investigate the influence of the 
ring exchange on the doped ladder, see Ref.\ 
\onlinecite{schmi05b} and references therein. 
A typical value is $x_{\square}\approx0.2$ \cite{nunne02, schmi05b}.
As for the examination of the anisotropic hopping we take the reliable 
results for $x=0$, $\lambda_{\perp} = \lambda_{\parallel} = 0.5$; $x=0.5$, 
$\lambda_{\perp} = \lambda_{\parallel} = 0.25$ and $x=1$, $\lambda_{\perp} = 
\lambda_{\parallel} = 2$ without ring exchange as starting point for our 
investigation.

For the first case $x=0$, $\lambda_{\perp} = \lambda_{\parallel} = 0.5$ the 
influence of the ring exchange is the weakest (see Fig.\ 
\ref{x=0 lambda=0.5 ring exchange}). The odd band hardly changes. It is 
slightly lowered -- only the maximum is decreased more strongly. For 
the even band, however, the changes are more pronounced. 
It is strongly lowered so that a 
band crossing occurs and the shape changes. A local miminum appears at $k=0$ 
developing into the global minimum with growing $x_{\square}$.

Fig.\ \ref{x=0.5 lambda=0.25 ring exchange} depicts the case 
$x=0.5$, $\lambda_{\perp} = \lambda_{\parallel} = 0.25$. Both bands are 
lowered stronger for small $k$ than for large $k$. Also the shape of both 
bands changes. While the minimum of the even band moves to $k\approx0.65\pi$ 
for $x_{\square}=0.2$ and a local maximum occurs at $k=\pi$, the local maximum 
at $k=0$ of the odd band moves to $k\approx0.55\pi$ and a local minimum at 
occurs $k=0$.

For $\lambda_{\perp} = \lambda_{\parallel} = 2$ and $x=1$ only the gs results 
are discussed because the pc generator yields no conclusive results for these 
parameters without ring exchange. The resulting one-hole dispersions are shown 
in Fig.\ \ref{x=1 lambda=2 ring exchange}. The increase of $x_{\square}$ yields
 a lowered dispersion for both bands. The shape of the bands is conserved, but 
for the odd band the decrease is pronounced around $k=\pi$. 
Even an increase of the energy can be observed 
for the even band around $k=\pi$.

\section{Summary}
\label{chap:summ}

The aim of the present paper was to establish an approach for doped
low-dimensional quantum antiferromagnets which yields an effective
model for the motion of charges. To this end, we employed a
self-similar continuous unitary transformation (SCUT) to lightly 
doped spin ladders. The method is particularly suited to
provide effective models beyond  numerical results for special
quantities. Spin ladders are a particularly well suited system
because their magnetic excitations are very well understood 
\cite{schmi05b,bouil11}.

We are able to calculate the one-hole dispersions by means of SCUT. The 
agreement with results from series expansion is very good for small parameters.
 Even in the regime $x=1$, $x_{\square}=0$, $2<\lambda<3$ the agreement with 
the exact diagonalisation results is still very good for the even band and 
satisfactory for the odd one. The hole dispersions are strongly influenced by 
the triplons. Because the one-hole dispersion with odd parity has a larger 
local energy which is influenced more strongly by the higher lying 
continuum, it changes more explicitly with increasing hopping constants.  
Therefore the deviations 
from the simple cosine shape are more pronounced for the odd band. These 
deviations grow if either the magnetic coupling or the strength of the hopping 
is increased:
\begin{itemize}
\item The broadening of the odd band is slowing down, then turning into a 
narrowing. Finally the shape changes completely under the influence of the 
second harmonic so that the maximum at $k=0$ is only local, the total maximum 
occurs at $k=\pi$ and the minimum lies between $0$ and $\pi$.
\item The shift upwards is also slowing down and then turning into a shift 
downwards due to the influence of the continuum formed by one triplon and 
one even hole state. In the regime around $x=1$ and $\lambda_{\perp} = 
\lambda_{\parallel} = 2$ the average odd dispersion is approximately 
as low as the average even dispersion.
\end{itemize}
For the even band the deviations from the cosine shaped dispersion consist 
essentially in the growth of the second harmonic. For $x=1$ and 
$\lambda_{\perp} = \lambda_{\parallel} = 2$ the minimum moves from $k=\pi$ 
into the region $k\approx\frac{\pi}{2}$. A local maximum at $k=\pi$ occurs. 
But the absolute maximum remains at $k=0$.

The combination of these effects for $x=1$ and $\lambda_{\perp} = 
\lambda_{\parallel} = 2$ yields a crossing of the two dispersions. 
The crossing point lies between the minima of the bands. However, the 
pc generator is not applicable for the SCUT if the parameters are in this 
region. This is suggested by the deviations from the series expansion and from
exact diagonalisation as well as by the peculiar convergence behavior of 
the SCUT.

In the regime $x\approx1$ and $\lambda_{\perp} = \lambda_{\parallel} \gtrsim 1$
 the pc generator is no longer suitable because the convergence of the flow is 
hindered by the overlap between the one-hole-one-triplon continuua and the 
one-hole-two-triplon continua. The pc results deviate from the series 
expansion and exact diagonalisation results. Furthermore the convergence is 
very slow and exhibits features that indicate problems in the sorting 
of the eigen values.

The remedy is to use 
the gs generator which only decouples the zero-triplon subspace 
from the remaining Hilbert space. 
Then the hole dispersions are similar to the exact 
diagonalisation results. Although there are small deviations from the series 
expansion, the agreement is astonishingly good. Because the exact 
diagonalisation studies a ladder with fourteen rungs, finite size effects are 
present so that a part of the deviations stem from them.
 The remaining deviations are probably due to truncation errors
. Another aspect in favour of the gs generator is the 
satisfactory convergence behavior.

The treatment of the case $\lambda_{\perp} =\lambda_{\parallel} > 2$ 
is problematic even for the gs generator because the convergence 
is hindered by the 
overlap between the odd one-hole dispersion and the continuum formed by one 
even hole and one triplon. To achieve convergence in this regime we have 
developed the following restriction for the gs generator. A term affecting the 
hole-triplon continuum is omitted, if the distance between the triplon and the 
hole state on which the term acts is larger than $\Delta n_{\mathrm{max}}$. 
Strictly, the effective Hamiltonian yields an upper boundary for the hole 
dispersions. The comparison of the results from the full gs generator and 
from the restricted 
gs generator for $x=1$ and $\lambda_{\perp} = \lambda_{\parallel} = 2$ shows 
that the upper boundary given by the result from the restricted generator is 
already very close to the result from the full generator for 
$\Delta n_{\mathrm{max}}=2$. For $x=1$ and $\lambda_{\perp} = 
\lambda_{\parallel} = 3$ this restriction  induces 
convergence, while the flow diverges for $\Delta n_{\mathrm{max}}>2$. The 
estimates we obtain for the hole dispersions agree again well with the exact 
diagonalisation results.

If $x=1, x_{\square}=0$ and
 $\lambda=\lambda_{\perp} =\lambda_{\parallel} = 3$ or 
$\lambda=\lambda_{\perp} =\lambda_{\parallel} = 2$  the bands exhibit a 
similar shape and relative position to each other so that still a crossing at 
$k\approx0.5\pi$ occurs. But the energy is lowered and the bandwith of both 
bands is increased by a factor of $\approx1.5$ if $\lambda$
is changed from 2 to 3. The results from the 
restricted gs generator yield an estimate for the even band which is in good 
agreement with the exact diagonalisation result, while the odd hole dispersion 
exhibits deviations that are very similar to the deviations between the exact 
diagonalisation and the full gs generator for $\lambda_{\perp} =
\lambda_{\parallel} = 2$.

The case $x=\lambda_{\perp}=\lambda_{\parallel}=0.5$ is a special one. Both the
 pc and the gs generator (even with increased maximal extensions) exhibit 
deviations from the series expansion results for the odd hole state. These 
deviations stem apparently from the closeness (or even overlap) of the 
continuum formed by one even hole and one triplon. If the lowering of the odd 
band is overestimated by the SCUT or underestimated by the series expansion 
is not clear.

The ring exchange leads to a lowering of the hole dispersions. This effect is 
least pronounced for the even band around $k=\pi$. The deformation of the 
band shape is most pronounced for the odd band.

Summarizing, the essential achievement of this paper is
the development of a unitary transformation  which
allows us to disentangle the motion of the doped charges, the holes, and
the magnetic excitations, triplons. Thereby an effective
model for the motion of the holes and the triplons has been derived.
The challenge was to systematically derive this effective model
even in the experimentally relevant regime $\lambda\approx3$ where
the hopping takes about three times the value of the magnetic
exchange constants. This challenge could only be met by modification
of the unitary transformation in the spirit of what
was done in Ref.\ \onlinecite{fisch10a}.

\section{Outlook}
\label{chap:outlook}

Ladders doped with more than one hole have not yet been 
treated by the SCUT. To do so the approach presented here
has to be extended to include also interaction terms between
two holes. This step will be subject of future research
because it will address the key question how large the
attractive forces between two doped charges are.
In view of the Cooper pair formation in doped
cuprate superconductors this is a very interesting issue.
Of course, numerical and diagrammatic results for bound
states of two holes exist \cite{white97,jecke98,jurec02,roux05}.
But so far no effective model has been derived which
incorporates the attractive potential explicitly.

The fact that we could realize a systematic mapping
of the initial doped spin ladder to an effective model
for the motion of single excitations
is encouraging for the next step incorporating the hole-hole
interaction. The additionally required resources, for instance
the increase in the number of coefficients, are significant.
We estimate that about six times more terms have to be 
kept track of. Still this should be realizable in the 
near future.

We expect that a substantial gain in the understanding
of Cooper pair formation and thus of superconductivity
in doped Mott insulators will be possible.

\begin{acknowledgments}
We are indebted to A.M. L\"auchli for the provision of the
exact diagonalization data shown in Figs.\ \ref{ED AL x=1 lambda=1},
\ref{ED AL x=1 lambda=2}, and \ref {ED AL x=1 lambda=3}. 
We thank Kai P. Schmidt, Carsten Raas, Tim Fischer, Simone A. Hamerla and Nils 
Drescher for fruitful discussions. The initial stage of this project was 
supported by the Graduiertenkolleg ``Materials and Concepts for Quantum 
Information Processing'' (GK726) funded by the Deutsche Forschungsgemeinschaft.
\end{acknowledgments}

\appendix

\section{Finite Size Scaling of the Exact Diagonalisation Results}

\label{app:fin_size_scal}

\begin{figure}
\begin{center}
\includegraphics[scale=0.3]{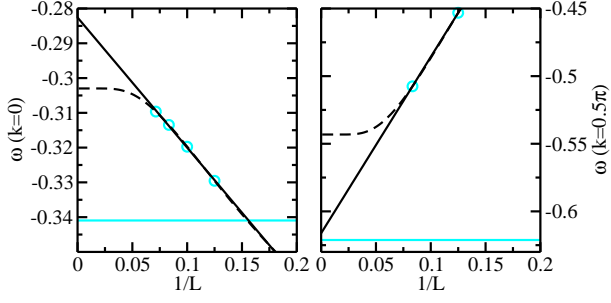}
\caption{(Color online) Finite size scaling of the exact diagonalization 
results for $x=1$ $\lambda=1$ in case of the odd band at $k=0$ and at 
$k=0.5\pi$. The cyan (gray) circles depict the exact diagonalization results 
for finite ladders with various numbers of rungs $L$. The solid black line is 
a linear extrapolation with respect to $\frac{1}{L}$ and the dashed line is an 
extrapolation based on exponential saturation (see Eq.\ 
(\ref{exponential saturation})). For comparison the result from the gs SCUT is
 also shown (solid cyan/gray).}
\label{finite size scaling x=1 lambda=1}
\end{center}
\end{figure}

To understand the deviations between SCUT and exact diagonalization in case of 
the odd band we also study the finite size scaling of the exact 
diagonalization. First we consider the exact diagonalization results for $x=1$ 
and $\lambda=1$ shown in Fig.\ \ref{ED AL x=1 lambda=1}. Around 
$k\approx0.25\pi$ our result for the odd band lies above the exact 
diagonalization result, while it lies below the exact diagonalization result 
around $k\approx0.6\pi$.

An investigation of the finite size scaling for the 
exact diagonalization data in these regions is difficult because we have two 
points at most for an extrapolation. Thus we consider $k=0$, where an 
extrapolation is conclusive. The result of this extrapolation is used to 
support an extrapolation in the region where the deviations are observed. The 
left panel of Fig.\ \ref{finite size scaling x=1 lambda=1} shows the finite 
size scaling for $k=0$ using two kinds of extrapolation. The first is a simple 
linear extrapolation with respect to $\frac{1}{L}$, while the second assumes 
an exponential saturation with increasing $L$ so that
\begin{equation}
\label{exponential saturation}
\Delta\omega \propto \mathrm{e}^{-\frac{L}{\xi}}
\end{equation}
holds true for the difference $\Delta\omega$ from the limit for 
$L\rightarrow\infty$. The correlation length $\xi$ is determined to be 
$\approx4.32$ by this extrapolation. The correlation length can be used to 
apply the second extrapolation also for $k=0.5\pi$ where only two results are 
obtained by exact diagonalization. 

The finite size scaling for $k=0.5\pi$ in 
case of the odd band is investigated by both extrapolations in the right panel 
of Fig.\ \ref{finite size scaling x=1 lambda=1}. The extraploation results at 
$k=0$ are still close to the gs SCUT result and the extrapolation results at 
$k=0.5\pi$ are in good agreement with the gs SCUT result. The linear 
extrapolation is even in excellent agreement with our result. Note that the 
points from the $L=6$ calculation are omitted for the extrapolations because 
they deviate from the behavior of the other points due to the very small 
size.

\begin{figure}[t]
\begin{center}
\includegraphics[scale=0.3]{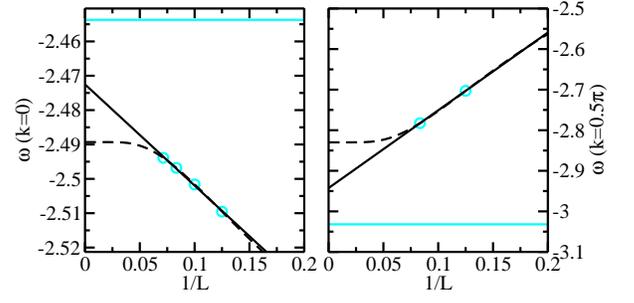}
\caption{(Color online) Finite size scaling of the exact diagonalization 
results for $x=1$ $\lambda=2$ in case of the odd band at $k=0$ and at 
$k=0.5\pi$. The cyan (gray) circles depict the exact diagonalization results 
for finite ladders with various numbers of rungs $L$. The solid black line is 
a linear extrapolation with respect to $\frac{1}{L}$ and the dashed line is an 
extrapolation based on exponential saturation (see Eq.\ 
(\ref{exponential saturation})). For comparison the result from the gs SCUT is 
also shown (solid cyan/gray).}
\label{finite size scaling x=1 lambda=2}
\end{center}
\end{figure}

For $x=1$ and $\lambda=2$ the SCUT result for the dispersions comes close to 
the extrapolation of exact diagonalization results in the regions where the 
continuum is distinguishable from the hole dispersion. This is most obvious 
for the odd band at $k=0$. The left panel of 
Fig.\ \ref{finite size scaling x=1 lambda=2} shows the finite size scaling 
for this case using the same 
extrapolations like for $x=1$ and $\lambda=1$. We determine $\xi$ to be 
$\approx4.03$ by the extrapolation based on exponential saturation. The 
correlation length is again used to apply this kind of extrapolation also for 
$k=0.5\pi$. 

The finite size scaling for $k=0.5\pi$ in case of the odd band is 
investigated by both extrapolations in the right panel of Fig.\ 
\ref{finite size scaling x=1 lambda=2}. For both momenta the linear 
extrapolation comes close to the SCUT result, but also the extrapolation 
based on exponential saturation does not exhibit large deviations from 
the SCUT result. The deviation for $k=0$ is $\approx2\%$ and the deviation 
for $k=0.5\pi$ is $\approx7\%$. The points from the $L=6$ calculation are 
omitted for the extrapolations because they deviate from the behavior of 
the other points due to the small size of the system.

For $x=1$ and $\lambda=3$ the tendency of the finite size scaling at 
$k=0.5\pi$ indicates that the exact diagonalization results overestimate the 
energy of the odd band in this region so that one can expect that a proper 
finite size scaling yields a dispersion that lies completely below the upper 
boundary provided by the restricted SCUT energies.

 However, an extrapolation from the two points ($L=8$ and $L=12$) at 
$k=0.5\pi$ is not conclusive. We have several points for an extrapolation for 
$k=0$ and $k=\pi$, but there the distinction between continuum and dispersion 
is difficult. Because the gs SCUT diverges without restriction of the 
generator, we actually expect that a strong overlap  
of the energies of single holes  and of the energies of hole
plus triplon states is present.


\end{document}